\documentclass[
    ,final            
  ]
  {aipproc}

\layoutstyle{8x11double}

\usepackage{makeidx}         
\usepackage{graphicx}        
\usepackage{multicol}        
\usepackage{url}             
\usepackage[bottom]{footmisc}
\usepackage{float, amsmath, natbib, epsfig}

\begin{document}

\title{Filamentation Instability of Interacting Current Sheets 
in Striped Relativistic Winds: The Origin of Low Sigma?}


\author{Jonathan Arons}{
  address={Astronomy Department, Physics Department and Theoretical Astrophysics Center \\
  University of California, Berkeley}
}

\begin{abstract}
I outline a mechanism, akin to Weibel instabilities of interpenetrating beams, in which the neighboring current sheets in a striped wind from an
oblique rotator interact through a two stream-like mechanism (a Weibel instability in 
flatland), to create an anomalous
resistivity that heats the sheets and causes the magnetic field to diffusively annihilate
in the wind upstream of the termination shock. The heating has consequences for observable unpulsed emission from pulsars.
\end{abstract}

\maketitle

\section{Introduction \label{sec:intro}}

 Observations over the last 15 years \cite{gaens06, slane05} and dynamical models \cite{bogo05, delzanna04, delzanna06, komiss03,  spit04} have confirmed the early suggestions \cite{emmer87, ken84, rees74} that PWNe behave as if the MHD wind from the underlying pulsars are weakly magnetized as the plasma emerges from the wind's termination shock (TS). If some form of shock acceleration underlies the conversion of the upstream flow's energy flux to the observed nonthermal spectra of synchrotron emitting electrons and positrons, with particle energies from 10s of MeV to PeV, the MHD shock jump conditions imply that $\sigma_1 = B_1^2 /4\pi \rho_1 \gamma_1 c^2$, the magnetization just upstream of the shock, must be substantially less than unity.  Studies of the flow dynamics in the nebulae, especially just outside of the termination shocks, arrive at the same conclusion.  The most advanced comparisons of dynamical theory (both MHD and beyond MHD) to date suggest that the latitude averaged upstream $\sigma$ in the Crab Nebula is $\sim 0.02$ \cite{delzanna06, spit04}. 
 
 Low $\sigma$, and, from energy conservation, high wind Lorentz factor $\Gamma_w$,
 are puzzles, since elementary ideal MHD applied to relativistic wind outflows suggest
 small wind acceleration and $\sigma \gg 1$ in the asymptotic wind. As a result,
 there are several interpretations of the low $\sigma$ conclusion, depending on precisely what one means by the shock transition to which the jump conditions are applied.  Begelman \cite{begel99} suggested that the shock really is a jump in a high $\sigma$ flow, but the toroidal magnetic field is MHD unstable with respect to 3D kink motions, whose result is to introduce downstream magnetic dissipation that rapidly coverts magetic energy into heat, hopefully in the form of the observed nonthermal particle spectra.  If this layer is thin, then the shock jump conditions apply to the overall layer, including both the true shock and the downstream dissipative layer.  The viability of this idea requires 3D MHD modeling, which has yet to appear; from an observational perspective, kink instability driven motions which can lead to actual dissipation also tangle the field, which creates the risk that the polarization of the synchrotron emission in the final model is less than that observed \cite{veron93}.
 
Lyubarsky \cite{lyubarsky03}, drawing upon a specific model of  the striped upstream wind of the oblique rotator \cite{lyubarsky01} which suggested slow dissipation of the magnetic field and its current sheets, has suggested that the wind arrives at the shock with $\sigma_1 \gg 1$ but appears as if it is low sigma with respect to the downstream nebula because magnetic dissipation (driven by reconnection) occurs within the shock front itself \cite{lyubarsky03, petri07} - the wavelength of the stripes is much less than the Larmor radii of the $e^\pm$ in the shock's magnetic field, which facilitates rapid dissipation.  Coroniti \cite{coroniti90} originally suggested that the current sheets of the striped wind disipate rapidly in the upstream wind, so that the MHD shock really does form in a $\sigma_1 \ll 1$ flow.  As pointed out in \cite{kirk03}, for causality reasons this can happen only if the flow has four velocity $c \beta_{w1} \Gamma_{w1}$ just upstream of the TS that is relatively low compared to the early estimate $\Gamma_{w1} \sim 10^6 $ obtained in \cite{ken84, rees74} - these early models considered a wind whose plasma content is inadequate to model the radio emission from PWNe, where the inefficiency of synchrotron emission from lower energy particles requires a wind with much larger mass loading than is required to explain the emission of optical and higher energy photons from the nebulae. 

I reconsider Coroniti's model, by introducing a new model for a two-stream like filamentation instability of the curent sheets in the striped wind that draws its free energy from the interaction between neighboring sheets.  The instability has close kinship with the Weibel instability that occurs between interpenetrating plasma streams in an unmagnetized, homogeneous plasma. I use a simple magnetic trapping model of the instability's saturation to estimate the anomalous resistivity that appears within the sheets, and I find the dissipation rate to be closely related to Bohm diffusion.  The resulting dissipation is rapid, with the sheet's broadening speed exceeding $0.1c$, which leads to annihilation of the striped magnetic field within a broad equatorial sector of the outflow in the upstream wind,  thus offering a viable model for the origin of low sigma in 
 the plasma arriving at the TS. I also suggest
 the radiation from the heated upstream plasma is the origin of the unpulsed optical emission observed in the Crab pulsar \cite{kanbach05}.
 
 \section{Plasma Outflow and Sheet Beam Injection \label{sec:injection}}

Modern theory suggests pulsars' rotational energy loss, occurring at the rate
$\dot{E}_R = 4 \pi^2 I \dot{P}/P^3$, appears as a relativistic MHD wind. The particle injection rates into the young PWNe require these flows' compositions to be largely electrons and positrons, with 
$\dot{N}_\pm \gg   \dot{N}_{GJ} =  c\Phi/e = 10^{34} (\Phi /10^{16.6} \; {\rm V}) $
s$^{-1}$,  the Goldreich-Julian charge loss rate, with 
$\Phi = (\dot{E}_R /c)^{1/2} = 4 \times 10^{16} [(\dot{P}/10^{-12.35})
(33 \; {\rm msec}/P)^3]^{1/2}$ V the magnetospheric voltage.  Magnetospheric models
show that the angular momentum flow is maintained by a combination of conduction current and displacement current.   In the model considered here, the dissipation of those currents in the equatorial sector $|\lambda| < i$ of the wind zone, $R_L =cP/2\pi =48,000P \; {\rm km} \ll r \ll R_{TS} (\approx 10^9 R_L$ in the
Crab), where $i$ is the angle between the neutron star's magnetic moment and its 
angular velocity, with $\lambda$  the latitude with respect to the rotational equator, is the origin of the weak asymptotic magnetization. The conduction currents have magnitude $I = c\Phi$, and form a 
system of primary and return currents, as revealed by recent force free and  MHD
models of relativistic magnetospheres. 
Synchrotron models of PWNe photon emission 
require substantial outflows of electron-positron plasma - for the Crab Nebula, 
$\kappa_\pm \equiv \dot{N}_\pm /   \dot{N}_{GJ} > 10^6$, a value well in excess
of the pair multiplicities of {\it outflows} found in extant polar gap and outer gap models - see \cite{arons08} for a brief review of these.  The magnetization of the wind at launch is
$\sigma (R_L) = \sigma_0 /\Gamma_{wL} h_{wL} $, where $\Gamma_{wL}$ is the wind's bulk flow Lorentz factor at the light cylinder, $h_{wL}$ is the proper relativistic enthalpy per particle in units of $mc^2$ at the same radius,  
$\sigma_0 = \Omega^2 \Psi^2 / \dot{M} c^3 $, $\Omega$ is the star's angular frequency, $\Psi = R_L \Phi$ is the open magnetic flux per sterradian and 
$ \dot{M} = 2 m_\pm \kappa_\pm \dot{N}_{GJ} = 2 m_\pm  \kappa_\pm c\Phi/e 
(=  2.4 \times 10^{-13} (10^6/\kappa_\pm) \; M_\odot /{\rm yr}$ in the Crab) is the rest mass loss rate.   Reasonable extrapolation of the low frequency radio
spectrum in the Crab suggests $\kappa_\pm \sim 10^7$!
Then $\sigma_0 = e\Phi/2 \kappa_\pm m_\pm c^2 = 3.9 \times 10^3 (10^7 /\kappa_\pm)$ in the Crab]. 
If the wind neither accelerates nor heats, $\sigma$ is conserved in the outflow.
Energy conservation suggests that $\Gamma_{wL} h_{wL } $ might be similar to that found in existing pair creation models, which have $\Gamma_{wL} h_{wL} \sim 100$  \cite{hibsch01, harding02}. If so,  
$\sigma_L = \sigma (r = R_L) \sim 200 $ 
for the Crab's wind\footnote{For the Crab pulsar, the value $\sigma_L   \sim 10^4$
is often quoted. This figure uses the Kennel and Coroniti evaluation of the wind's properties, using their  simplified MHD model  which ignores the particle flux required to account for the radio emission.}.  Such a large value of $\sigma_0$ indicates the wind is very magnetically dominated, and would have an
asymptotic 4 velocity $\Gamma_{w\infty}^{ideal} \approx \sigma_0^{1/3} ( \sim 34$
in the Crab), if there is no magnetic dissipation in the wind zone. 

The nebular MHD models behave as if $\sigma \ll 1$ at the TS.  IF such weak magnetization is a property of the {\it upstream} wind, then the pulsar's energy flux is 
carried by the particles at the termination radius.  Within the MHD wind model, energy 
conservation tells us that the terminal value of the 4 velocity is simply 
$\Gamma_{w\infty} = \dot{E}_R /\dot{M} c^2 = e \Phi /2 \kappa_{\pm} m_\pm c^2 = \sigma_0 $

\subsection{Lessons from the Aligned Rotator}

\begin{figure}[h]
  \includegraphics[height=.3\textheight]{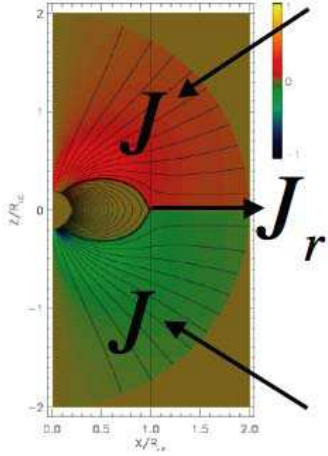}
  \caption{Magnetic field and electric current flow structure of the aligned rotator
  $i = 0$), from \cite{spit06}.  The poloidal field is almost dipolar well inside the light cylinder, and approaches the split monopole for $r \gg R_L$.  The color shows the opposite senses of the toroidal magnetic field in the opposite hemispheres. The volume currents in the wind correspond to electron outflow, shown by the arrows marked $J$.  The return current $J_r$, corresponding to outflowing ions from the stellar surface plus outflowing positrons, probably extracted from the wind on open field lines, flows mostly in the current sheet. \label{fig:aligned}}
\end{figure}

Michel \cite{michel73} derived the structure of the magnetic field and Poynting flux component  of the force free split monopole's outflow. The current sheet separating the hemispheres
lies in the rotational equator, the magnetic field has poloidal components (in spherical
coordinates $B_r = \pm \Phi R_L/r^2, ; B_\theta = 0$ and toroidal field 
$B_\phi = \mp (\Phi /r) \cos \lambda$).  The monopole has strength $\mu /R_L$, 
where $\mu$ is the dipole moment - thus the magntiude of the monopole's field equals the equatorial strength of 
the static dipole at the light cylinder distance $R_L$. The electric field is 
$\vec{ E}  = - (\vec{ \Omega} \times \vec{r} ) \times
  \vec{B} = \mp (\Phi /r) \cos \lambda \hat{\boldsymbol \theta}$, which yields the energy 
  (Poynting) flux 
  $F = (\dot{E}_R /4\pi r^2) \cos^2 \lambda, \; \dot{E}_R = I\Phi = c \Phi^2$.  The signs are for the aligned case $i = \angle ({\vec\Omega}, \vec{\mu}) =0$.
The anti-aligned rotator $i = \pi$ has the signs reversed.  Figure \ref{fig:aligned} shows
the fields and currents of this system.

\begin{figure}[h]
 \includegraphics*[width=3in]{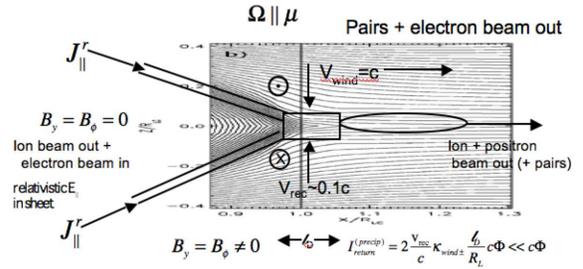}
 \caption{ A schematic model of reconnection at the Y-line for the aligned rotator. Contours show the poloidal magnetic field $B_p$; the circled cross and dot show the directions of the toroidal field. The electron-positron wind launched from the polar cap flows out on the open field lines with speed $c\beta_w \sim c$. A reconnection electric field (not shown) gives the plasma a velocity across $B_p$ with speed $v_{rec} \sim 0.1 c$ \cite{bessh05}, injecting $e^\pm$ into the diffusion region outlined by the rectangle, whose length is
  $l_D \sim c/\omega_p \ll R_L$, and into the closed zone.  A parallel electric field along the separatrix accelerates injected electrons toward the star, forming a component of the return current $I_r^{(e)} = 2(v_{rec} /c) \kappa_\pm (l_D / R_L) c \Phi \ll c \Phi $. Positrons precipitating along the separatrix are repelled and provide a contribution to the return current in the equatorial current sheet with the same magnitude. The rest of the return current is ions attracted up from the surface by the precipitating electrons \cite{arons83}.
 \label{fig:reconn}}
\end{figure}

The formation of these currents depends on physics not contained in the force-free model.  Reconnection at the equatorial Y-line probably plays a central role in the plasma dynamics, as is revealed in rudimentary manner in the evolutionary 
numerical solutions by \cite{komiss06, spit06}, where numerical resistivity allows sporadic, time dependent reconnection to occur. Analytical modeling of the sporadic formation of X-lines
and equatorial plasmoids emerging along the current sheet (Arons, in preparation) 
shows that flux transfer events from the open to the closed field lines, and back, inject pair plasma from the wind into a sheath surrounding the separatrix, that the electric field along the separatrix sends electrons down toward the boundary of the polar cap and positrons out along the equatorial current sheet, but that these fluxes are insufficient to provide the whole return current - the precipitating electrons, which have number density in excess of the Goldreich-Julian density, support the electric field required to pull ions up from the surface (they are Goldreich and Julian's ``hanging charge clouds") and launch the ions and the positrons extracted from the wind  into the current sheet beyond the light cylinder.  Figure \ref{fig:reconn} shows the structure of this model for the case 
$i = 0$. The anti-aligned rotator ($i = \pi$) has the same structure, with positrons extracted from the
wind precipitating onto the polar cap forming part of the return current, electrons extracted from the star forming the rest of the return current from the stellar surface and these electrons plus those extracted by reconnection from the wind forming the return current in the current sheet, while the volume current in the wind is ions extracted directly from the star's polar cap.

\begin{figure}[h]
 \includegraphics*[width=3in, trim=0in 0.75in 1in 3.75in]{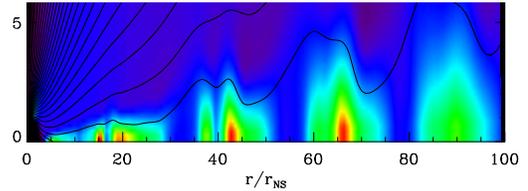}
 \caption{ Plasmoids formed by sporadic reconnection in the equatorial current sheet of the relativistic aligned rotator, from \cite{bucc06}. Blue represents unreconnected field, while yellow and red show poloidal field reshaped into O type rings.  These calculations, done for a very rapidly rotating proto-neutron star, have a light cylinder at $ r \sim 7$ neutron star radii. The plasmoids recede at the speed of light, and correspond to $\sim 20$\% fluctuations in the Poynting flux, with a fluctuation time on the order of the rotation period.
 \label{fig:plasmoids}}
\end{figure}

The current injected into the equatorial current sheet therefore is a beam (ions + positrons when $i = 0$, electrons when $i = \pi$), and the current sheet is a transmission line. In detail, the injector probably operates spasmodically, as is illustrated in Figure \ref{fig:plasmoids}. 

\subsection{Oblique Rotators - Striped Winds}

\begin{figure}[h]
\includegraphics[width=3in]{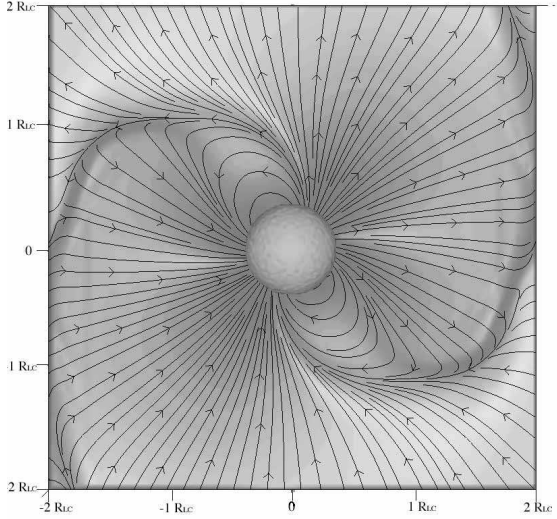}
 \caption{Poloidal field lines and cross section of the total currents in the 
  $ \Omega, \mu $ 
 plane for the $ i = \pi /3$ acute rotator, from \cite{spit06}. 
 Note the topological similarity to the aligned rotator
 \label{fig:oblique}}
\end{figure}

Observed pulsars are oblique rotators, $ 0 < i < \pi$.  In terms of the current flow 
structure, ``acute rotators'' ($i < \pi /2$) are like the aligned rotator, while ``obtuse
rotators'' ($i > \pi/2$) are like the anti-aligned rotator. The structure is illustrated in
Figure \ref{fig:oblique}.

The current sheet is now twisted, and takes on the form of a frozen in wave advected
with the outflow, which is well represented in Bogovalov's \cite{bogo99} analytic model 
of the
asymptotic wind, modeled as an oblique split monopole  - the meridional cross section of the current sheet in the simulations and in that simple analytic model, shown in Figure \ref{fig:meridian}, have very similar form:
\begin{figure}[h]
\includegraphics[width=3in]{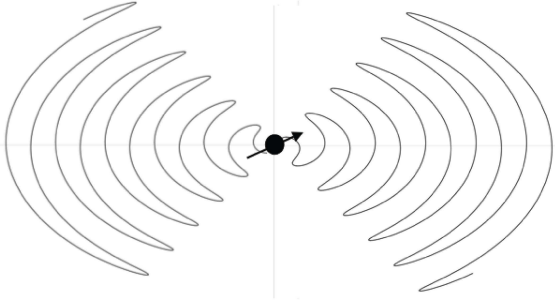}
 \caption{Meridional cross section of the total currents in the 
  $ \Omega, \mu $ 
 plane for the $ i = \pi /3$ oblique split monopole, from \cite{bogo99}. 
 \label{fig:meridian}}
\end{figure}
The field in the polar regions is circularly polarized, forming a helical wind, while in the equatorial region with latitude $|\lambda| < i \; (|\lambda| < \pi -i $ for the obtuse rotator), the magnetic field
is in the form of oppositely directed toroidal stripes, each stripe's B field having the strength expected from the split monopole $B = \pm (\Phi /R_L r)\cos \lambda$ and with full wavelength 2$R_L$. Pairs of stripes are separated by the current sheet where the beams flow, with the surface current between each pair being $j = \pm cB/2\pi $. If the current in the sheets dissipates, the toroidal field in the sectors disappears, creating an 
unmagnetized (or weakly magnetized) equatorial sector in the outflow, just what appears to be the case in PWN dynamics. Possible dissipation mechanisms are a) some form of anomalous resistivity forms in the sheets, causing them to heat and broaden and merge, destroying the magnetic stripes through magnetic diffusion \cite{coroniti90}; b) reconnection launched by tearing or  drift kink instabilities
(\cite{komiss07, zenitani07} and references therein), whose flows cause conversion of magnetic field to chains of regions with O topology strung along the sheets, which merge and destroy the striped field as the hot plasma injected from each X point expands; and c) mode conversion, in which the sheet currents vary coherently and convert the frozen in current sheet and the interleaved magnetic stripes into a relativistically strong electromagnetic wave propagating in an unmagnetized plasma, whose eventual dissipation yields an unmagnetized equatorial flow (\cite{skar03} and references therein).  I explore mechanism a) here, since it offers the possibility of there being a surviving weak component of the toroidal field, which is the simplest explanation of the ordered polarization observed near the termination shocks of the PWNe. 

\section{A Causality Limit \label{sec:causality}}

No matter what the mechanism of dissipation, there is a causality limit on whether
upstream dissipation can ever annihilate the stripes. All the mechanisms act as if the current sheets broaden at some speed $v_{spread} <c$.  Until dissipation occurs,  the wrinkled current sheet is frozen into the wind, flowing out at 4 velocity $c \beta_w \Gamma_w$. The sheets' separation in the wind's proper frame is $x_0 = \Gamma_w R_L$. The proper time for the sheets to merge and the stripes to dissipate is then
$x_0 /v_{spread} = \Gamma_w R_L /v_{spread}$. The dissipation time in the ``lab'' frame, where the neutron star's center of mass is at rest, is longer by a factor of 
$\Gamma_w: \; T_{diss} =   \Gamma_w^2 R_L /v_{spread}$.  Meanwhile, the flow time for a stripe to go from neutron star to the TS is $T_{flow} = R_{TS}/c\beta_w$. More or less complete stripe dissipation can occur only if $T_{diss} < T_{TS}$, which requires 
$\Gamma_w < \sqrt{(R_{TS}/R_L ) (v_{spread}/c) }$. For the well studied Crab 
Nebula, this upper limit on the wind 4-velocity is 
$\Gamma_w < 5.6 \times 10^4 (v_{spread} /c)^{1/2}$. This upper limit exceeds the 
terminal wind 4 velocity 
$\Gamma_{w\infty} = \sigma_0= 3.9 \times 10^3 (10^7 /\kappa_\pm)$ if 
$v_{spread} / c > 0.005 (10^7 /\kappa_\pm )^2 $.  
 
 \section{Filamentation Instability of Interacting Sheet Currents
  \label{sec:filaments} }

\begin{figure}[h]
 \includegraphics*[width=3in, height=2in, trim= 1in 1in 1in 1in]{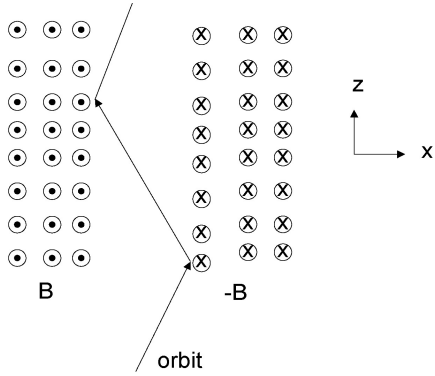}
 \caption{ Schematic orbit of beam particles in the unmagnetized center of the
 current sheet.  The external magnetic field of the stripes confines the sheet particles. These orbits are consistent with the dynamical behavior of the sheet
 considered as a fluid - see expression (\ref{eq:sheetdynamics}). \label{fig:channel}}
\end{figure}

As outlined above, the current sheets are beams injected into the wind,
transported to large radius in the unmagetized layer between the stripes.  Figure
\ref{fig:channel} illustrates the essence of the structure. The unmagnetized layer has half
thickness equal to the formal Larmor radius of the beam 
$H = r_{Lb} = m_b c^2 \gamma_b \beta_b /q_b B_0$, where the beam has 4 velocity
$c \beta_b \gamma_b$ and surface number density 
$\Sigma_b = 2 n_b H = j/cq_b \beta_b$, where $j= cB_0/2\pi$ is the surface current 
required to flip the magnetic field from $B_0 $ in a stripe to $-B_0$ in the stripe's neighbor. A simplified model of the sheet represents its internal structure as a 
region where $B = 0$ with thickness $2H$  occupying $-H \leq x \leq H$, extending
infinitely in $y, z$; the current within the sheet flows along the $z$ axis.
The particles scatter from the magnetic walls with angles $\psi$, creating a 
pressure of the beam on the magnetic field which holds the walls apart, 
$P_b = (\Sigma_b /2H ) \gamma_b \beta_b^2 m_b c^2 \langle \sin^2 \psi \rangle
\equiv (\Sigma_b /2H) T_{b\perp} = B_0^2 /8 \pi$.  Use of the beam model 
and $H = r_g$ in the jump condition $j= cB/2\pi$ then yields 
$T_{b\perp} = m_b c^2 \gamma_b \beta_b^2 /2$ and therefore 
$\langle \sin^2 \psi \rangle = 1/2$. 

\begin{figure}[h]
 \includegraphics*[width=3in, height=2in]{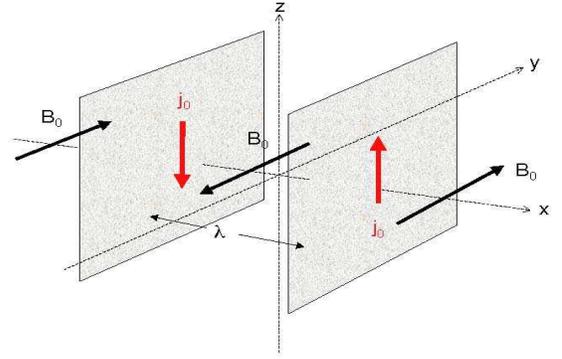}
 \caption{ A pair of current sheets, separating three stripes in the equatorial wind. The 
 $x$ direction corresponds to the radius, the toroidal magnetic field  
 $\pm B_0 = \pm \Phi /\Gamma_w r$ lies along the $y$ axes, and the sheet currents
 $\pm cB_0 /2\pi $ are in the $z$ direction.
 \label{fig:sheets}}
\end{figure}

Applying the same procedures as are 
used to describe a dynamically passive discontinuity in a plasma \cite{stix92} yields the equation of motion of a thin current sheet separating oppositely directed, equal magnetic fields - the plasma within the current sheet has dynamics as if the background magnetic field is absent:
\begin{eqnarray}
m_b c \Sigma_b  \frac{D(\gamma_b  \vec{\beta}_b ) }{Dt}  &  =  &
   q_b \Sigma_b (\langle \vec{E} \rangle + \vec{\beta}_b \times \vec{B}) - 
     \vec{\nabla}_\perp P_b \label{eq:sheetdynamics}  \\
     & = &
    q_b \Sigma_b (\langle \delta \vec{E} \rangle + \vec{\beta}_b \times \delta \vec{B} )
      - \vec{\nabla}_\perp T_b \delta \Sigma_b, \nonumber
\end{eqnarray}
where I have used $ \langle \vec{B}_0 \rangle= 0$. Here $\langle \rangle$ is the average
of the contained quantity's values on each side of the sheet, with the variable evaluated
in the surrounding intersheet medium.  That medium bwteen the sheets (the stripes) is well represented by ideal MHD.
Figure \ref{fig:sheets} shows a pair of sheets, separated by the distance 
$x_0 = \Gamma_w R_L$.

\begin{figure}[h]
 \includegraphics*[width=3in, height=2in]{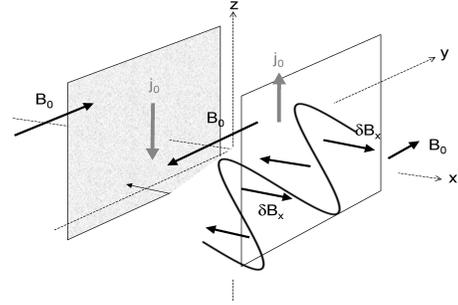}
 \caption{ Alfvenic disturbance of a pair of current sheets. The intersheet medium
 supports the waves as MHD disturbances ($B_0^2 / 4\pi \rho_0 c^2$ can be arbitrary,
 but is very large in the pulsar wind application.) \label{fig:alfven}}
\end{figure}

\begin{figure}[h]
 \includegraphics*[width=3in, height=2in]{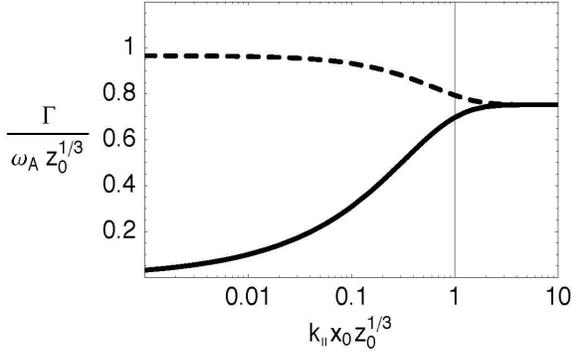}
 \caption{ Proper growth rate of filamentary magnetic fields for two interacting current sheets. Here 
 $\omega_A = k_\parallel v_A, \; 
 v_A =c\beta_A \left (1 + \frac{4\pi \rho_0 c^2}{B_0^2} \right)^{-1/2}, \; \rho_0 =$ 
 proper rest mass density of the wind plasma in a stripe between the sheets, $B_0 =$ proper stripe magnetic field, $z_0 =  \beta_A^2 \beta_b^2 /k_\parallel H$.  The solid
 line is the relevant branch. 
 I employ this result with $k_\parallel x_0 \sim 1$. \label{fig:dispersion}}
\end{figure}

Now imagine an electromagnetic disturbance of this structure with Alfven wave polarization $\delta B_x (x,y) = \delta \hat{B}_x (x) \exp [i(k_\parallel y - \omega t)]$, as shown in Figure \ref{fig:alfven}.
The $\vec{j}_0 \times \langle \delta \vec{B}_x \rangle $ force points parallel to the direction of the background magnetic fields and compresses each sheet's surface density into filaments
with axes parallel to the original current flow direction. These surface current filaments reinforce the original $\delta B_x$ and cause it to grow - a Weibel instability in flatland.
The dispersion relation including the sheet separation is surprisingly simple
\cite{arons07}, and yields
the approximate result 
\begin{equation}
\omega = i\Gamma_{2sheet} =
        i \frac{c}{x_0} \beta_A (\beta_b \beta_A k_\parallel x_0)^{2/3} 
            \left( \frac{x_0}{H} \right)^{1/3}.
 \label{eq:growth}
 \end{equation}
 The full solution of the dispersion relation appears in Figure \ref{fig:dispersion}.
In the initial cold wind leaving the light cylinder with 
$\Gamma_w \approx (e\Phi /2 m_\pm c^2 \kappa_\pm)^{1/3}$, the growth time (as 
measured in the neutron star's frame) becomes shorter than the expansion time scale of 
a fluid element when 
$r/R_L \gg (e\Phi /2 m_\pm c^2 \kappa_\pm)^{2/3} (\gamma_{b,lab} m_b /
2 \kappa_\pm m_\pm )^{1/2} \sim 
10^{3.3}  (\gamma_{b,lab} m_b /2 \kappa_\pm m_\pm )^{1/2} \ll R_{TS}/R_L \sim 10^9$;
the numerical values are for the Crab. Thus the current sheets develop sustained 
Weibel-like turbulence inside the sheets in the deep inner wind ($r \ll R_{TS}$). 
So long as the
striped magnetic fields persist, they pinch the sheets, forcing the current to flow, thus
driving the Weibel turbulence, which creates a ``collisional'' drag on the current carrying 
particles in the sheets.

Long analytical and simulation experience with Weibel instabilities in quasi-homogeneous 
media (\cite{chang07} and references therein) suggests
that this instability saturates by stochastic trapping of beam particles in magnetic potential wells formed by the magnetic fluctuations (equivalent to the current density in the filaments reaching the Alfven critical current.) That result, which implies turbulence amplitude  $\delta B$ such that 
$\delta \omega_c = q_b \delta B/ m_b c \gamma_b \approx \Gamma_{2sheet}$
 has not yet been demonstrated for the inhomogenous, interacting but separated layer driven instability
discussed here, but is a quite likely outcome.  Assuming trapping as the saturation mechanism, one readily finds an effective collision frequency 
$\nu_c = \langle (\delta \omega_c )^2 \rangle \tau_{ac} 
= \Gamma_{2sheet} (\Gamma_{2sheet} \tau_{ac} ) \equiv K_c \Gamma_{2sheet},
K_c \geq 1$; if the autocorrelation time of a particle in the turbulence is long,
$K_c \gg 1$ is possible. The conductivity inside the sheet is 
$\sigma_{b} = \omega_{pb}^2 /4 \pi \nu_c$, from which one readily finds
the magnetic diffusivity
\begin{equation}
D_b = \frac{c^2}{4\pi \sigma_b} = \frac{1}{3} cH \alpha_{b} 
   \left( \frac{H}{x_0} \right)^{2/3},
\label{eq:Bohm}
\end{equation}
with $\alpha_{b} \equiv   \beta_b \beta_A K_c 
       (k_\parallel x_0 \beta_b \beta_A)^{2/3} \approx K_c \geq 1$.  Since $H = r_g$,
       the formal gyration radius in the stripe fields bounding each sheet, 
(\ref{eq:Bohm}) suggests this instability leads to a variant of Bohm diffusion for the
magnetic diffusivity.

The scattering introduces Ohmic heating into the sheets through the formation of a resistive electric field in the proper frame of the flow 
$E_z = J_{beam}/\sigma_{b} = j_b/2H \sigma_b $. The internal energy per particle of the beam $e_b = T_{b\perp}/(\hat{\gamma} -1)$ changes non-adiabatically according to the 
comoving energy equation for one-dimensional thickening of the sheet 
$\dot{e}_b + (\hat{\gamma} -1)(\dot{H} / H) e_b = J_b E_z /n_b 
= (D_b /H) q_b \beta_b B_0  
= (\hat{\gamma} -1)\alpha_b  q_b c \beta_b B_0 (H/x_0)^{2/3}$. $\hat{\gamma}$ is the ratio
of specific heats of the beam as it undergoes magnetic scattering, 
$3/2 \geq \hat{\gamma} \geq 4/3$.
 Since $H = 2 T_{b\perp}/q_b \beta_b B_0$, when 
 $ \beta_b = $ constant $\approx 1$, the energy equation becomes
\begin{equation}
\frac{1}{B_0} \frac{d(B_0 H)}{dt}  + (\hat{\gamma} -1) \frac{dH}{dt} = 
 (\hat{\gamma} -1) \alpha_b c \left (\frac{H}{\Gamma_w R_L } \right)^{2/3},
 \label{eq:energy}
 \end{equation}
 with $B_0 = \Phi /\Gamma_w r$ until the late phases of current sheet expansion,
 when $D_b$ approaches the full Bohm rate as the sheets broaden to consume
the stripes ($r_g = H \rightarrow x_0$) and the stripe's $B$ field drops below the ideal MHD value.

Heating implies pressure forces that can accelerate the wind, causing $\Gamma_w$ to increase with increasing $r$ \cite{lyubarsky01}. The most pessimistic
estimate of the efficacy of this resistive model for sheet dissipation appears if one assumes $\Gamma_w$ is a constant with value equal its maximum 
$\dot{E}_R /\dot{M} c^2 = e\Phi /2 \kappa_\pm m_\pm c^2$. 
Then (\ref{eq:energy}) yields 
$H / \Gamma_w R_L = \left[ (\hat{\gamma} -1) \alpha_b /\beta_b \Gamma_w^2 \right]^3
(r/R_L)^3$, which implies sheet dissipation is complete ($H = \Gamma_w R_L $)
at $r = R_{merge} = \Gamma_w^2 R_L /(\hat{\gamma} -1) \alpha_b < R_{TS}$
if $\Gamma_w = \Gamma_{w\infty} \dot{E}_R/\dot{M}c^2 = 3.9 \times 10^3 (10^7 /\kappa_\pm) < 3 \times 10^4 \sqrt{2\alpha_b /3} $ (Crab). Also, if the autocorrelation time of the trapping is long and the resistivity is large, $K_c \gg 1$ and 
therefore $\alpha_b \propto K_c \gg 1$; a PIC simulation of the instability is needed in order to evaluate this possibility.  Thus the stripe dissipation radius is probably 
at $R_{merge} \sim 0.1 T_{TS} $, if not smaller, at least in the Crab pulsar's wind.   

\section{Conclusions}

I have argued that due to unstable interaction between the beam currents in the 
{\it neighboring} sheets that separate the magnetic stripes in the oblique rotator's 
wind, anomalous resistivity develops in the sheets which causes them to heat, 
expand and
consume the magnetic field of the stripes well upstream of the termination shock of
the Crab pulsar's wind, and quite likely in the winds of other pulsars.   The dissipation mechanism put forward here falls in
Kirk \& Skj{\ae}raasen's \cite{kirk03} ``fast'' category, with the sheet broadening velocity approaching c/3 toward the end of the process. Contrary to the conclusion of Lyubarsky and Kirk \cite{lyubarsky01}, who constructed a model
in Kirk \& Skj{\ae}raasen's ``slow'' category, I find that dissipation of the magnetic 
field in the equatorial, striped zone can and probably does occur in the flow 
upstream of the termination shock, as was suggested by Coroniti \cite{coroniti90}.  More complex dissipative flows, such as relativistic tearing and drift-kink instabilities \cite{zenitani07} would only enhance 
this conclusion - estimates indicate that simple thickening of the sheets dominates 
over these more complex flows once the sheets have substantially broadened, although they may play a role in the early non-linear dynamics of the sheets.

My conclusion
rests on the winds being heavily mass loaded, as is indicated by radio observations of PWNe, therefore having asymptotic wind 4-velocities much less \cite{gallant02} than the estimates in \cite{ken84, rees74}, who neglected the implications of the radio emission in 
their pioneering modeling efforts. Theoretically, such large mass loading is not 
understood. Theoretical models of pair creation in pulsar magnetospheres with any pretense of self-consistency all underpredict pulsars' mass loss rates by one, two or 
more orders of magnitude. Combined with the basic conflict between 
the poloidal currents found in force-free models 
of the underlying magnetospheres with the currents implied by the extant pair 
creation models (both ``polar cap''
and ``outer gap''), these problems suggest a substantial rethinking of pair creation 
physics and wind formation is in order \cite{arons08}. 

Most of the dissipation happens far from the star, where $\Gamma_w$ is close to
to its maximum. Because of relativistic beaming, radiation from the inefficiently 
emitting, resistively heated plasma would appear as a steady point source 
superposed on the pulsar.  Optical, polarimetric observations of the Crab pulsar have amply demonstrated
the existence of unpulsed emission with flux $\sim$ 1-2\% of the pulse peak intensity 
\cite{kanbach05}.  Preliminary estimates indicate that synchrotron emission from
the resistively heated beams and the pair plasma  from the stripes engorged by the expanding sheets
may well be the origin of this emission.  If so, the radiatively dark winds upstream of their termination in the surrounding nebulae may be subject to observational 
investigation.

\begin{theacknowledgments}
 
 During the slow and continuing unfolding of this project, I have particularly 
 benefitted
 from discussions with S. Cowley, F. Coroniti and B. Schmekel.  My research efforts on this and other topics have been supported by
NSF grant AST-0507813,  NASA grant NNG06G108G and DOE grant DE-FC02-06ER41453, all to the University of California, Berkeley; by the Department of Energy contract to the Stanford Linear Accelerator Center no. DE-AC3-76SF00515; and by the taxpayers of California.

\end{theacknowledgments}





\begin{thebibliography}{99}

\bibitem[Arons(1983)]{arons83}
Arons, J. 1983a, in {\it Positron-Electron Pairs in Astrophysics} (New York: AIP), 163

\bibitem[Arons(2007)]{arons07}
Arons, J. 2007, submitted to ApJ

\bibitem[Arons(2008)]{arons08}
Arons, J. 2008, ``Pulsars: Progress, Problems and Prospects'',  in Springer Lecture Notes on "Neutron Stars and Pulsars, 40 years after the discovery", ed. W.Becker,
in press (astro-ph/0708.1050)

\bibitem[Begelman(1999)]{begel99}
Begelman, M. 1999, ApJ, 512,755

\bibitem[Bessho \& Bhattacharjee(2005)]{bessh05}
Bessho, N., and Bhattacharjee, A. 2005, Phys. Rev. Lett., 95, 245001

\bibitem[Bogovalov(1999)]{bogo99}
Bogovalov, S.V. 1999, A \& A, 349, 1017

\bibitem[Bogovalov {\it et al.}(2005)]{bogo05}
Bogovalov, S.V., Chechetkin, V.M., Koldoba, A.V., and Ustyugova, G.V. 2005,
  MNRAS, 358, 705
  
\bibitem[Bucciantini {\it et al.}(2006)]{bucc06}
Bucciantini, N., Thompson, T., Arons, J., {\it et al.} 2006, MNRAS, 368, 1717
  
 \bibitem[Chang, Spitkovsky \& Arons(2007)]{chang07}
Chang, P., Spitkovsky, A., and Arons, J. 2007, submitted to ApJ (astro-ph/0704.3832)
  
\bibitem[Coroniti(1990)]{coroniti90}
Coroniti, F. 1990, ApJ, 349, 538

\bibitem[Del Zanna {\it et al.}(2004)]{delzanna04}
Del Zanna, L., Amato, E. and Bucciantini, N. 2004, A \& A, 421, 397

\bibitem[Del Zanna {\it et al.}(2006)]{delzanna06}
Del Zanna, L., Volpi, D. Amato, E. and Bucciantini, N. 2006, A \& A, 453, 621

\bibitem[Emmering \& Chevalier(1987)]{emmer87}
Emmering, R.T., \& Chevalier, R. 1987, APJ, 321, 334

\bibitem[Gallant {\it et al.}(2002)]{gallant02}
Gallant, Y.A., van der Swaluw, E., Kirk, J.G., and Achterberg, A. 2002, in {\it Neutron Stars
  in Supernova Remnants}, P.O. SLane and B.M. Gaensler, eds. (San Francisco: ASP
  Conference Series Vol. 271), 99

\bibitem[Gaensler \& Slane(2006)]{gaens06}
Gaensler, B., and Slane, P. 2006, Ann. Rev. Astro. Astrophys., 44, 17

\bibitem[Harding {\it et al.}(2002)]{harding02}
Harding, A.K., Muslimov, A.G., and Zhang, B. 2002, ApJ, 576, 366

\bibitem[Hibschman \& Arons(2001)]{hibsch01}
Hibschman, J.A., and Arons, J. 2001c, ApJ, 560, 871

\bibitem[Kanbach {\it et al.}(2005)]{kanbach05}
Kanbach, G., Slowikowska, A., Kellner, S., and Stenle, H. 2005, in {\it Astrophysical Sources of High
 Energy Particles and Radiation} (New York: AIP Conference Proceedings), 306

\bibitem[Kennel \& Coroniti(1984)]{ken84}
Kennel, C.F., and Coroniti, F.V. 1984, ApJ, 283, 694; 283, 710

\bibitem[Kirk \& Skj{\ae}raasen(2003)]{kirk03}
Kirk, J.G., and Skj{\ae}raasen, O. 2003, ApJ, 591, 366

\bibitem[Komissarov \& Lyubarsky(2003, 2004)]{komiss03}
Komissarov, S., and Lyubarsky, Y. 2003, MNRAS, 344, L93; 2004, {\it ibid.}, 349, 779

\bibitem[Komissarov(2006)]{komiss06}
Komissarov, S. 2006, MNRAS, 367, 19

\bibitem[Komissarov {\it et al.}(2007)]{komiss07}
Komissarov, S., Barkov, M. and Lyutikov, M. 2007, MNRAS, 374, 415

\bibitem[Lyubarsky \& Kirk(2001)]{lyubarsky01}
Lyubarsky, Y., \& Kirk, J. 2001, ApJ, 547, 437

\bibitem[Lyubarsky(2003, 2005)]{lyubarsky03}
Lyubarsky, Y. 2003, MNRAS, 345, 153; 2005, Adv. Space Res., 35, 1112

\bibitem[Michel(1973)]{michel73}
Michel, F.C. 1973, ApJ, 180, L133

\bibitem[Petri \& Lyubarsky(2007)]{petri07}
Petri, J., \& Lyubarsky, Y. 2007, submitted to A\&A (astro-ph/0707:1782)

\bibitem[Rees \& Gunn(1974)]{rees74}
Rees, M.J., and Gunn, J.E. 1974, MNRAS, 167,1

\bibitem[Skj{\ae}aasen {\it et al.}(2003)]{skar03}
Skj{\ae}aasen, O., Melatos, A., and Spitkovsky, A. 2003, ApJ, 634, 542

\bibitem[Slane(2005)]{slane05}
Slane, P. 2005, Adv. Space Res., 35, 1092

\bibitem[Spitkovsky \& Arons(2004)]{spit04}
Spitkovsky, A., and Arons, J. 2004, ApJ

\bibitem[Spitkovsky(2006)]{spit06}
Spitkovsky, A. 2006, ApJ., 648, L51

\bibitem[Stix(1992)]{stix92}
Stix, T. 1992, ``Waves in Plasmas'', (New York:AIP), 108-112

\bibitem[Veron-Cetty \& Woltjer(1993)]{veron93}
Veron-Cetty, M.P., \& Woltjer, L. 1993, A\&A, 270, 370

\bibitem[Zenitani \& Hoshino(2007)]{zenitani07}
Zenitani, S., \& Hoshino, M. 2001, ApJ, 562, L56; {\it ibid.}, 2007, ApJ, submitted (archiv:0708.1000) 

\end{thebibliography}

\end{document}